\documentstyle[11pt,fleqn,epsf]{article}
\def\ref{par\noindent\hangindent=6mm\hangafter=1}
\begin{document}
\vbox{
}
\baselineskip 8mm

\begin{center}
{\bf Noisy kink in microtubules}

\bigskip

H.C. Rosu$^{a,c}$
\footnote{Electronic mail:
rosu@ifug.ugto.mx},
J.A. Tuszy\'nski$^{b}$
\footnote{Electronic-mail:jtus@phys.ualberta.ca} 
and  A. Gonz\'alez$^{a}$
\footnote{Electronic mail:
gonzalez@ifug.ugto.mx}

$^{a}$
{\it Instituto de F\'{\i}sica de la Universidad de Guanajuato, Apdo Postal
E-143, L\'eon, Gto, M\'exico}

$^{b}$
{\it Department of Physics, University of Alberta, Edmonton,
AB, T6G 2J1, Canada}

$^{c}$
{\it Institute of Gravitation and Space Sciences,
Magurele-Bucharest, Romania}

\end{center}

\bigskip
\bigskip


ABSTRACT.
We study the power spectrum of a class of noise effects generated by means of 
a digital-like disorder in the traveling
variable of the
conjectured Ginzburg-Landau-Montroll kink excitations moving along the walls
 of the
microtubules. We have found a 1/f$^{\alpha}$ noise with $\alpha \in$
 (1.82-2.04)
 on the time scales we have considered.



PACS:  87.15.-v, 72.70+m, 87.10.+e \hfill

\newpage


One can find extensive, descriptive presentations of microtubules (MTs) 
in many biological papers.
Here, we shall give only elementary definitions as follows.
They are ubiquitous protein polymers in eukaryotic cells belonging to the
category of {\em biological filaments} and making
the most part of the cytoskeleton. They are hollow cylinders 25 nm in
outer diameter and 17 nm in the inner one, with lengths ranging from nm to mm
in some neuronal cells. The
walls of the cylinders are usually made of 13 (the seventh Fibonacci number)
protofilaments laterally associated.
The surface structure of MT walls is very interesting \cite{mand}.
Structurally, MTs are quasi one-dimensional chains of tubulin polar dimers
(negative $\alpha$ and positive $\beta$ monomers, each of 4 nm in length)
undergoing conformational changes induced by the guanosine triphosphate to
diphosphate (GTP-GDP) hydrolysis. The whole assembly process of a MT is due to
the hydrolyzation of GTP. The cation Mg$^{++}$ is
essential to increase the affinity of tubulin for binding GTP and thus for
generating MTs.
Moreover, a unique dynamical property is
the so-called {\em dynamical instability} \cite{-1} which is a random growth
and shrinkage of the more active plus ends of MTs. It has
raised much interest in recent years \cite{-2}.


An energy-transfer mechanism in MTs by means of Ginzburg-Landau-Montroll (GLM)
kinklike
protofilament excitations has been discussed by
Satari\'c, Tuszy\'nski and \u Zakula
\cite{1}
in 1993.
Also sine-Gordon (SG) solitons have been discussed by
Chou, Zhang and Maggiora in 1994
\cite{czm}.
We recall that various types of solitons have found interesting applications
in biological physics (Davydov's model \cite{dav}, DNA/RNA \cite{dna}).
Usually, in order to get
nonlinear differential equations one performs a continuum limit for some
lattice models in which discreteness effects are neglected. Some authors
have shown that such effects might be important and suggested
various ways of including them in continuous differential equations \cite{discr}.
The digital disorder we shall
comment on and use next is just a possible way to incorporate discreteness 
in a moving solitonic pattern.

In 1993, Rosu and Canessa
\cite{2} introduced a
digital-like disorder in the Davydov $\beta$-kink leading to the 1/f$^{\alpha}$
noise, with $\alpha\approx1$,
in the dynamics of that kink, and also commented on the multifractal features
of the dynamics of the $\beta$-kink. The procedure is as follows. In the
traveling variable $\xi=x-v_K t$ of the slowly moving kink (the acoustic
Lorentz factor $\gamma _L \approx 1$) one puts
$x_i=x_0+\Delta x_i$, where
$x_0$ is the position of the center of mass of the kink, or its central
position along the chain, and $\Delta x$ are small random displacements around
$x_0$ (say, in the interval $\pm 1$). In the calculations, one can fix
$x_0=0$. For each random position $\Delta x_i$ chosen from a uniform
distribution, one calculates by means of a fast-Fourier-transform algorithm
the {\em noise power spectrum} of the time series of the signal
(considered to be the kink) in order to get the time correlations of the
fluctuations of the signal, i.e.,
$$
S_{K}(f)\propto\frac{1}{\tau}\left\langle|\int_0^{\tau}K(\xi)e^{2\pi ift}dt|^2
\right\rangle ~,
\eqno(1)
$$
where $0<t<\tau\approx 1/f$ and K($\xi$) is the kink function, and 
the brackets stand for averaging over ensembles.
This approach to noise effects has been taken from the literature on
the self-organized criticality (SOC) paradigm, see, e.g., \cite{vm}.
On the other hand, the standard treatment of noise effects when they originate 
in thermal
fluctuations is by means of the Langevin equation method. Valls and Lust
\cite{vl} have studied the effect of thermal noise on the front propagation
in the GL case. They have found a crossover between constant-velocity
propagation at early times and diffusive behavior at late times.

The purpose of the present work is to investigate
the noise produced by the same type of disorder as in \cite{2}
in the case of the GLM kink conjectured in MTs.

The main assumption in \cite{1} is that the assembly of tubulin dimers/dipoles
(${\cal D}_n$) form a quasi one-dimensional ferrodistortive system
for which the double-well on-site potential model is a standard framework
$$
V({\cal D}_{n})=-\frac{1}{2}A {\cal D}_{n}^{2}+\frac{1}{4}B{\cal D}_{n}^{4}~.
\eqno(2)
$$
The variable
${\cal D}$ has been identified in \cite{1} 
with the amount of $\beta$-state distorsion vertically
projected
(the $\beta$-state is defined as having the mobile electron
within the $\beta$-monomer).
Moreover, in \cite{1} a GL hamiltonian/free-energy with intrinsic
electric field and dissipation effects included led to the dimer Euler-Lagrange
dimensionless equation of motion (EOM) in the
traveling coordinate of the anharmonic oscillator form (with linear friction)
$$
|A|{\cal D}^{''}-\gamma \alpha v_K{\cal D}^{'}
-F({\cal D})=0~,
\eqno(3)
$$
where $\gamma$ is the friction coefficient,
$\alpha=|A|\gamma ^2 _L/Mv_{sound}^2$ and
$F({\cal D})=A{\cal D}-B{\cal D} ^3+qE$,
with $q$ denoting the effective charge of a single dimer of mass M,
and $E$ the magnitude
of the intrinsic electric field. This EOM
is known to have a unique kink solution given by the formula
\cite{3}
$$
K (\xi)=a+\frac{b-a}{1+\exp (\beta \xi)}\equiv a+
\frac{\beta}{\sqrt{2}}(1-\tanh(\beta\xi /2))~,
\eqno(4)
$$
where $K={\cal D}/\sqrt{|A|/B}$ is a rescaled dipole variable,
$\beta=(b-a)/\sqrt{2}$, whereas $a$ and $b$ are two of the solutions
of the cubic equation
$$
F(K)\equiv(K -a)(K -b)(K -c)\equiv K^3 -K -\sigma=0~,
\eqno(5)
$$
where $\sigma =q\frac{\sqrt{B/|A|}}{|A|}E$
and $u_0$ units are used, where $u_0=\sqrt{|A|/B}\approx 1.4\cdot 10^{-11}$ m
is the amplitude of the dimer displacement (shift of the double-well potential).
Notice that the GLM kink is thin.
Its width is $w_K=\frac{1}{\beta}\approx 0.7 u_0$.

As we said, the type of digital disorder we consider here is very close
to the ideas
of the SOC paradigm that we understand in the
broad sense of {\em both} spatial and temporal scaling of the dynamical
state of the system \cite{soc}.
The spatial scaling (self-similarity) is of the (multi)fractal
type while the temporal scaling leads to $1/f^{\alpha}$ noises. 





A strong motivation for dealing with digital dynamics is the
possibility of generating broken symmetries and therefore of having various
types of dynamical phase transitions \cite{ch}.
Thus, digital disorder, though might look a rather ad-hoc approach,
focuses on both self-organized properties of MTs, i.e., to driven steady
states with long-range spatio-temporal correlations, and to (dynamical)
phase transitions, since digital dynamics allows for symmetry breaking.

Our results are displayed in Figures 1 and 2 and show that the noise introduced by 
the digital disorder in the GLM kink variable
is practically of the 1/f$^2$ (Brownian) type on the time scales 
we have considered.  
At low frequencies there is the known cross-over to a white noise due to the
finite system size, which is moving to lower frequencies as the size of the 
system is increased \cite{fj}. On the other hand, the deviation from the power
law at high frequencies is an artifact due to the so-called aliasing \cite{hjj}.
The Brownian noise we have obtained is not unexpected since it is 
a common occurence in SOC models in any finite dimension \cite{cc}, and only 
mean-field calculations reproduced the $1/f$ noise \cite{tb}.

Finally, we would like to mention that if polarization does not exactly follow 
the displacement one needs
a system of two coupled partial differential equations leading to
two coupled traveling kink waves.
Of course, the
nonlinear models seem to be too simple-minded for the MT complexity.
Nevertheless they provide guides for further insight and perhaps some
partial answers.
%

{\bf Acknowledgments}

This work was partially supported by CONACyT (Mexico) and by NSERC (Canada).



\newpage

\centerline{
\epsfxsize=250pt
\epsfbox{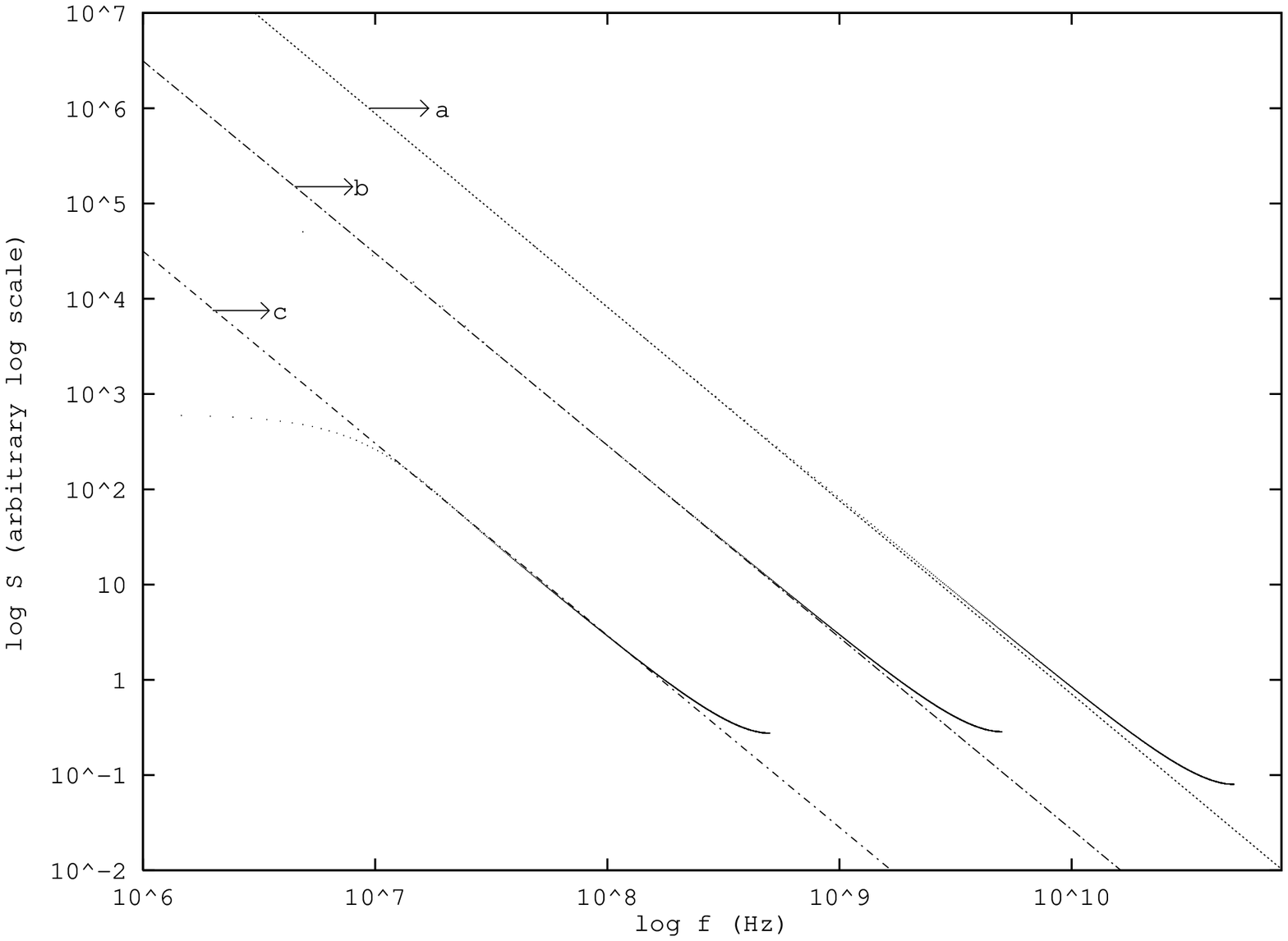}}
\vskip 4ex
\begin{center}
{\small{Fig. 1}\\
 Double logarithmic plot of the power spectrum of the digital noise 
perturbing the motion of the GLM kink moving along MTs at constant v$_K$
= 2m/s. The fitted slopes are as follows: (a) -2.032 (b) -2.0175 (c) -2.0164
for the time scales corresponding to that of the dimer, ten times bigger, and
hundred times bigger, respectively. The errors in the slopes are at the level
of 0.0003 for each case.}
\end{center}

\vskip 2ex
\centerline{
\epsfxsize=250pt
\epsfbox{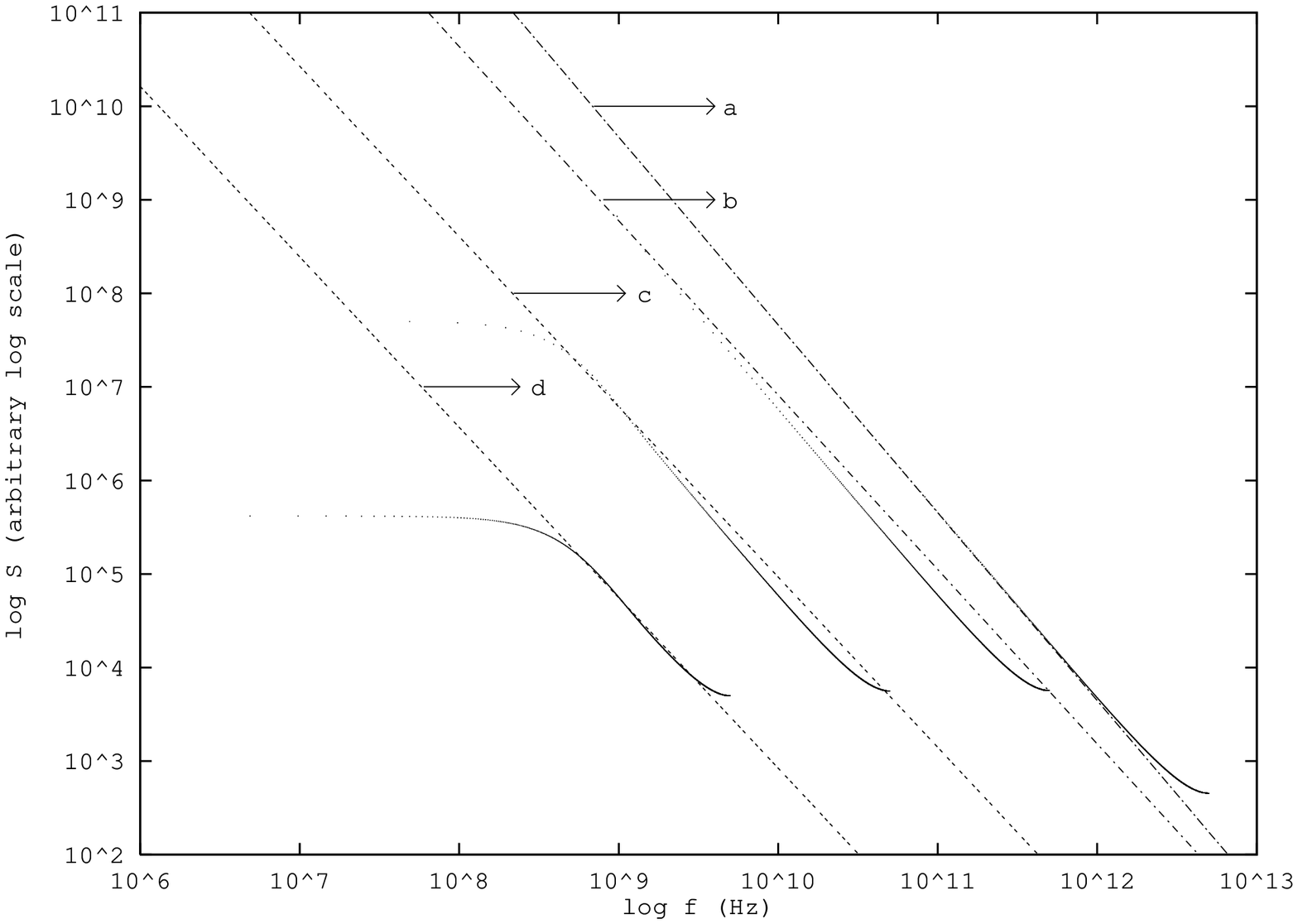}}
\vskip 4ex
\begin{center}
{\small{Fig. 2}\\
 The same plot as in Fig.~1 for  v$_K$
= 100m/s. The fitted slopes are (a) -2.0083 (b) -1.8641 (c) -1.8205 (d) -2.0000
for temporal scales of 10$^{-2}$, 10$^{-1}$, 10$^{0}$, and 10$^{1}$
times that of the tubulin dimer, respectively. The level of the errors is the 
same as in Fig.~1.}
\end{center}

\end{document}